%% file: main.tex
\documentclass[conference]{IEEEtran}
\IEEEoverridecommandlockouts
\usepackage{cite}
\usepackage{amsmath,amssymb,amsfonts}
\usepackage{algorithmic}
\usepackage{graphicx}
\usepackage{textcomp}
\usepackage{xcolor}
\usepackage{units}
\usepackage{multirow}
\usepackage{colortbl}
\usepackage{soul}
\usepackage[left=0.625in, right=0.625in,top=0.75in,bottom=1in]{geometry}

\usepackage{amsthm}
\usepackage{thmtools, thm-restate}
\usepackage{algorithm}
\input{symbols_commands}
\newcommand{\alg}{$\mathsf{ROAR-Fed}~$}

\allowdisplaybreaks[4]

\usepackage{hyperref}
\usepackage[hyphenbreaks]{breakurl}
\def\BibTeX{{\rm B\kern-.05em{\sc i\kern-.025em b}\kern-.08em
    T\kern-.1667em\lower.7ex\hbox{E}\kern-.125emX}}
\begin{document}

\title{RIS-Assisted Over-the-Air Adaptive Federated Learning with Noisy Downlink 
\thanks{This work is supported in part by NSF CNS-2112471.}
}

\author{Jiayu Mao and Aylin Yener
\\ INSPIRE@OhioState Research Center 
\\Dept. of Electrical and Computer Engineering
\\ The Ohio State University
\\ mao.518@osu.edu, yener@ece.osu.edu 
}

\newgeometry{left=0.625in, right=0.625in,top=0.75in,bottom=1in}

\maketitle

\input{Abstract}


\vspace{-0.05in}
\section{Introduction}
\label{sec:intro}
Over the last decade, federated learning (FL) \cite{mcmahan2017} has emerged as a promising distributed machine learning paradigm, and has attracted significant interest from both academia and industry.
FL exploits local computation capabilities and preserves the privacy of edge devices by coordinating the devices through a parameter server (PS) to collaboratively train a global model in an iterative manner.
Due to the time-varying nature of wireless channels and limited communication resources, care must be exercised in applying FL in wireless networks.

OTA-FL \cite{amiri2020} enables wireless federated learning by exploiting the innate superposition property of the wireless channels.
In particular, OTA-FL employs simultaneous analog communication of local update, and global update aggregation to occur naturally over the wireless medium.
Most prior works on OTA-FL have focused on uplink transmission with error-free downlink assumptions, while limited attention has been paid to the downlink-only noisy communication model.
Reference \cite{amiri2021convergence} proposes and analyzes downlink digital and analog transmissions assuming an error-free uplink.
Recent few works present convergence analyses of OTA-FL over uplink and downlink noisy communication channels and propose design schemes~\cite{wei2022federated,guo2022joint,qi2022robust,shah2022robust}.
However, the existing literature is dominated by system models that consider the availability of perfect channel state information (CSI) as accurate over-the-air model aggregation critically relies on CSI.
In practice, users can only obtain estimated CSI, which can lead to signal misalignment and degradation on the learning performance.
This paper considers noisy uplink and downlink communications under imperfect CSI at the edge devices that constitute an FL system.

Recently, reconfigurable intelligent surfaces (RIS) \cite{wu2019intel} have emerged as a novel technology for next-generation wireless networks such as 6G, by equipping a programmable meta-surface with massive low-cost passive reflecting elements and reconfiguring the propagation environment via independent phase shift adjustment of the incident signal.
The judicious deployment of RIS can proactively modify the channel conditions to create more favorable propagation environments, thereby it has the potential to facilitate the model aggregation stage when integrated in OTA-FL~\cite{yang2020fed}.
Several works have studied RIS-enhanced federated learning.
For example, \cite{zheng2022balancing}, \cite{ni2021fed} and \cite{liu2021csit} employ RIS to minimize the mean squared error (MSE) of model aggregation, while \cite{wang2021fed} considers the MSE as a constraint to maximize the number of participating devices.
\cite{li2022one} deploys RIS to aid a one-bit communication FL system.
More recently, unified learning and communication designs are proposed. 
In particular, \cite{liu2021risfl} develops a 
unified communication-learning optimization problem to jointly optimize RIS phase, beamforming and device selection, but assumes a static time-invariant channel with error-free downlink and perfect CSI.
\cite{zhao2022performance} proposes Lyapunov optimization method to minimize the optimality gap of RIS-aided FL and energy consumption.

In contrast with the majority of works that emphasize the minimization of MSE communication problem, in this work, we adopt the cross-layer approach that jointly optimizes the computation and communication resources simultaneously to enhance the learning performance in an RIS-assisted OTA-FL system.
We consider a practical time-varying physical layer with noisy uplink and downlink channels with imperfect CSI at the edge devices.
This model is nontrivial in the sense that the communication noise in uplink and downlink will accumulate during the iterative training procedure, while our previous work~\cite{mao22roar} only considers noisy uplink.
We assume a general FL setup with a non-convex objective function and a heterogeneous network with non-i.i.d. local data distribution and differing user resources.
We aim to provide a robust joint communication and learning algorithm with the aid of an RIS placed between the PS and the users.
Specifically, we design dynamic power control schemes on \emph{both downlink and uplink transmissions} in each global iteration, where we adapt local update steps, RIS phase matrix and transmit power in concert to combat the effects of both time-varying imperfect CSI and system/data heterogeneity.
We present the convergence analysis of the proposed algorithm and evaluate its performance with extreme non-i.i.d. local datasets.
Numerical results show that our approach achieves excellent test accuracy and outperforms the existing unified design under simultaneous downlink and uplink noisy channels and imperfect CSI, verifying the effectiveness and robustness of our algorithm for RIS-assisted OTA-FL system.

\section{System Model} \label{sec: prelim}
\vspace{-0.05in}
\subsection{Federated Learning Model} 
\label{subsec: fl}
We consider an FL model with $m$ users and a parameter server (PS).
User $i$ has a local dataset whose data points are sampled from the distribution $\mc{X}_i$.
In this work, we assume that users are heterogeneous in terms of their local datasets, as is the case in practice.
Specifically,  the distributions of local datasets are non-i.i.d., i.e., $\mc{X}_i \neq \mc{X}_j$ if $ i \neq j, \forall i, j \in [m]$.
FL aims to minimize a global empirical loss function:
\begin{equation}
    \min_{\w \in \mathbb{R}^d}F(\w) \triangleq \min_{\w\in\mathbb{R}^d} \sum_{i \in [m]} \alpha_i F_i(\w, D_i), 
    \label{eq: objective}
\end{equation}
where $\w$ represents the model parameter with $d$-dimension, $\alpha_i = \frac{| D_i |}{\sum_{i \in [m]} | D_i |}$ is the factor of user $i$, $F_i(\w, D_i) \triangleq \frac{1}{| D_i |} \sum_{\xi^i_j \in D_i} F(\w, \xi^i_j)$ denotes the local loss function, where $\xi^i_j$ is the $j$-th sampled data point. 
In this paper, we consider non-convex learning objective functions, i.e., $F_i(\w, D_i)$ is non-convex as in practice.
And users have different numbers of training data points, i.e., $\alpha_i \ne \alpha_j$ if $i\ne j$.

In FL, learning is performed through collaborative training between PS and users. 
In each communication round, users train the local models using their own dataset and send the updated model parameters to the PS. The PS then updates the global model accordingly from aggregated local models.
The next round starts with the server broadcasting the global model parameters, and the entire learning process iteratively continues until the global model converges.
Note that in OTA-FL, communication and aggregation happen at the same time at the PS due to the innate superposition property of the wireless channel.

Specifically, in the $t$-th round, the PS sends the global model parameter $\w_t$ to users. 
Next, each user $i$ starts computing the local gradient by its dataset $D_i$. 
The local training is performed through the stochastic gradient descent (SGD) method.
User $i$ starts with initialization $\w^i_{t, 0}$ and trains locally for $\tau_t^i$ steps:
\begin{equation}
    \w^i_{t, k+1} = \w^i_{t, k} - \eta_t \nabla F_{i}(\w^i_{t, k}, \xi^i_{t, k}), \quad k = 0,\ldots,\tau_t^i-1, \label{equ:sgd}
\end{equation}
where $\xi^i_{t, k}$ is a single random data point in the $k$-th local step.
The number of local steps $\tau_t^i$ varies across clients in every communication round, as in our previous works \cite{yang22,mao22,mao22roar}.
\vspace{-0.05in}
\subsection{RIS-Assisted Communication Model} 
\label{subsec: comm}
\vspace{-0.05in}
\subsubsection{Uplink Communication Model} 
\label{subsubsec: uplink}
\begin{figure}[t] 
    \centering
    \includegraphics[scale=0.25]{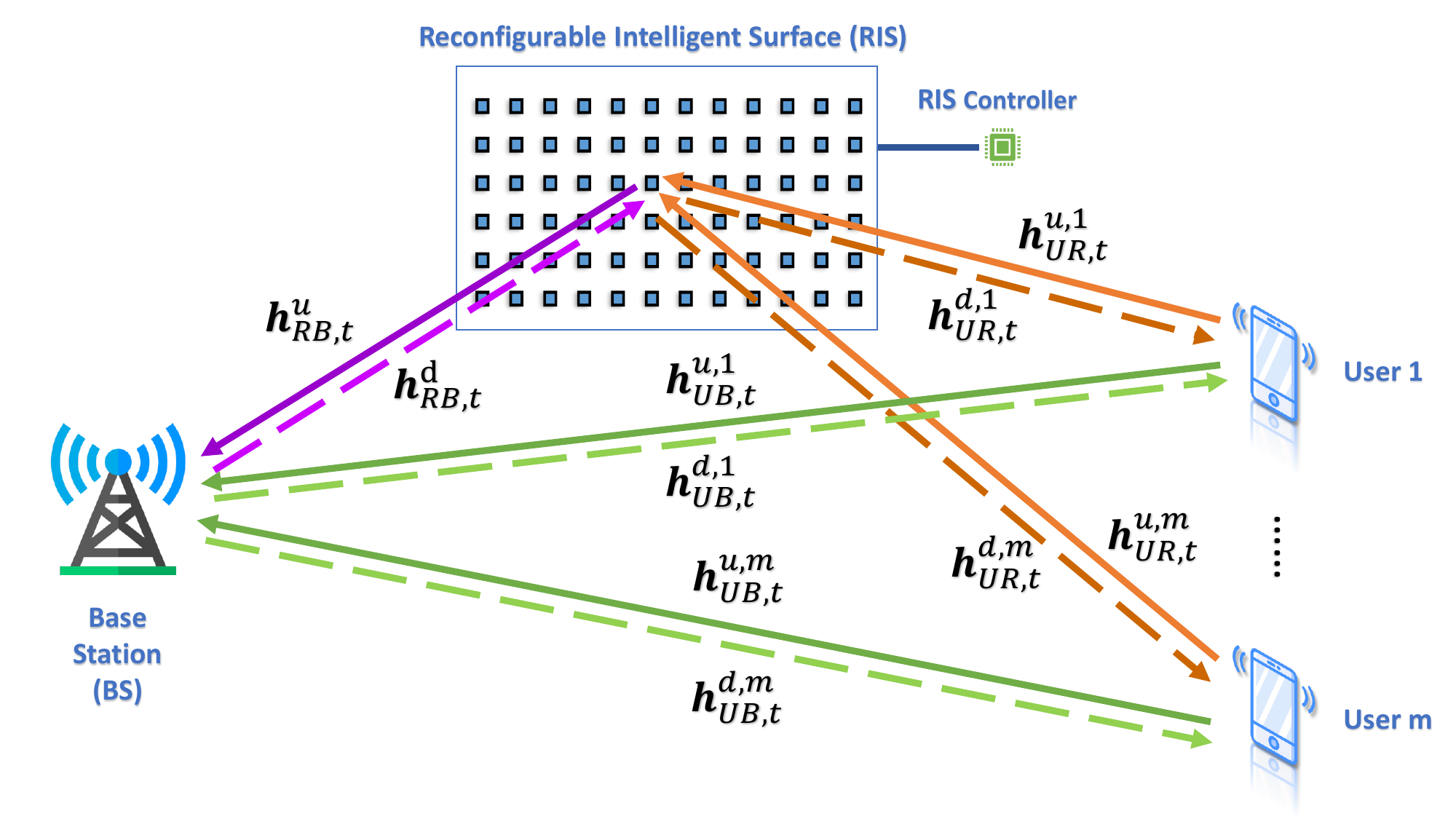}
    \caption{The RIS-assisted noisy uplink and downlink communication system.}
    \label{fig:sysmodel}
\end{figure}

We investigate an RIS-assisted uplink communication model \footnote{Without loss of generality, we consider synchronous models.} as in Fig.~\ref{fig:sysmodel}.
The system contains $m$ users and one base station/PS, all of which are equipped with a single antenna.
A single RIS with $N$ passive phase shifting elements is placed between the PS and edge devices.
We assume that the direct links are weak, hence the presence of RIS is significant.
We consider a block fading channel model, where the channel coefficients remain constant during each FL uploading stage and change independently in each global round.
Denote $h_{UB,t}^{u,i} \in \Cb$, $\h_{RB,t}^{u} \in \Cb^N$, $\h_{UR,t}^{u,i} \in \Cb^N$ as uplink channels from user $i$ to PS, from RIS to PS, from user $i$ to RIS in the $t$-th round, respectively.
We define a diagonal matrix $\The_t$ as RIS reflecting matrix, i.e., $\The_t = diag (\theta_{1,t}, \theta_{2,t},\cdots, \theta_{N,t})$, where $\theta_{n,t} = e^{j \phi_{n,t}}$ is the $n$-th continuous reflecting element.
To assist communication more efficiently, we design the RIS phase elements to update accordingly in each global round similar to~\cite{mao22roar}.
The corresponding received signal $\y_t^u$ at the BS is given by
\begin{equation}
    \y_t^u = \sum_{i \in [m]} (h_{UB,t}^{u,i} + (\h_{UR,t}^{u,i})^H \The_t \h_{RB,t}^u)\x^i_t + \z_t^u, \label{equ:upreceivesig}
\end{equation}
where $\x_t^i \in \mb{R}^d$ denotes the transmit signal from user $i$, $\z_t^u$ is an i.i.d. additive white Gaussian noise (AWGN) with zero mean and variance $\sigma_{c,u}^2$.
For ease of notation, we define the $i$-th RIS assisted link as $\g_t^{u,i} = ((\h_{UR,t}^{u,i})^H \Hb_{RB,t}^u)^H \in \Cb^N$, where $\Hb_{RB,t}^u = diag(\h_{RB,t}^u)$.
Then we denote the RIS reflecting vector as $\theb_t=(\theta_{1,t},...,\theta_{N,t})^T$.
As such, the received signal can be equivalently written as $\y_t^u = \sum_{i \in [m]} (h_{UB,t}^{u,i} +  (\g_t^{u,i})^H \theb_t)\x^i_t + \z_t^u.$
We also consider an individual power constraint for user $i$ in $t$-th communication round: $\mathbb{E}[\| \x^i_t \|^2] \leq P_t^i, \forall i \in [m], \forall t, $ where $P_t^i$ is the maximum transmit power.

\subsubsection{Downlink Communication Model} 
\label{subsubsec: downlink}
We consider an RIS-assisted downlink communication model as in Fig.~\ref{fig:sysmodel} in the FL broadcasting stage.
Similarly, we assume weak direct links and block fading channel model with variant channel gains in each communication round.
Denote $h_{UB,t}^{d,i} \in \Cb$, $\h_{RB,t}^{d} \in \Cb^N$, $\h_{UR,t}^{d,i} \in \Cb^N$ as downlink channels from PS to user $i$, from PS to RIS, from RIS to user $i$ in the $t$-th round, respectively.
Each user $i$ receives the broadcasting signal $\y_t^{d,i}$ as:
\begin{equation}
    \y_t^{d,i} = (h_{UB,t}^{d,i} + (\h_{RB,t}^d)^H \The_t \h_{UR,t}^{d,i})\x_t^d + \z_t^{d,i}, \label{equ:dlrecvsig}
\end{equation}
where $\x_t^d$ is the downlink transmit signal from PS, $\z_t^{d,i}$ is the i.i.d. downlink additive noise with zero mean and variance $\sigma_{c,d}^2$ at device $i$.
We consider a power constraint at PS: $\mathbb{E}[\| \x^d_t \|^2] \leq P_t^d, \forall t,$ where $P_t^d$ is the downlink power budget at the $t$-th round.

We assume that each edge device has access to imperfect channel state information (CSI) for both uplink and downlink transmissions. In other words, we assume that users have estimated CSI.
For simplicity, we use $\hh_t$ to represent the estimated CSI of each wireless link: $\hh_t = h_t + \Delta_t, \forall t,$
where $\Delta_t$ is the i.i.d. estimation error with zero mean and variance $\sigt_h^2$. All links have channel estimation error.
For ease of notation, we define the effective $i$-th-user-PS channel coefficient in $t$-th round as
\begin{equation}
    h_t^{u,i}=h_{UB,t}^{u,i} +  (\h_{UR,t}^{u,i})^H \The_t \h_{RB,t}^u,
\end{equation}
\begin{equation}
    h_t^{d,i}=h_{UB,t}^{d,i} + (\h_{RB,t}^d)^H \The_t \h_{UR,t}^{d,i}.
\end{equation}
Similarly, we use $\hh_t^{u,i}$ and $\hh_t^{d,i}$ to represent the overall estimated uplink and downlink CSI, respectively.
\vspace{-0.05in}
\section{Joint Communication and Learning Design} 
\label{sec: alg}

\begin{algorithm}[t!] 
    \caption{RIS-Assisted Over-the-Air Adaptive Federated Learning with Noisy Downlink} \label{alg:apaf} 
    \begin{algorithmic}[1]
    \STATE 
    \emph{\bf Initialization: $\w_0$, $\theb_0$, $\beta_t^i$, $\tau_t^i, i \in [m]$.}
    \FOR{$t=0, \dots, T-1$}
    \STATE{\bf $\bullet$ Phase update:}
    \STATE {PS selects the user with maximum $\tau_{t-1}^i$ and updates RIS phase shifts by SCA method~\eqref{equ:phase}.} 
    \STATE{\bf $\bullet$ Downlink transmission:}
        \STATE {PS broadcasts global model $\w_t$ by~\eqref{equ:dltransig}, users estimate the starting global model weights by~\eqref{equ:wt0i}.}
    \STATE{\bf $\bullet$ Uplink transmission:}
        \FOR{user $i, i \in [m]$ in parallel}
        \STATE {User $i$ computes $\tau_t^i$ to meet communication constraints and does local training. Then it finds $\beta_t^i$ by~\eqref{equ:betaiimp} and uploads $\x_t^i$.} 
        \ENDFOR
    \STATE {PS updates the global model by~\eqref{equ:globalup}.}
    \ENDFOR
    \end{algorithmic}
\end{algorithm}

In this section, we propose a cross-layer approach that simultaneously optimizes communication, computation, and learning resources.
Specifically, we propose a design where we update the RIS phase at the beginning of every global round and keep it fixed throughout the entire iteration. Then we design dynamic power control schemes for both downlink and uplink transmissions.

\vspace{-0.05in}
\subsection{Downlink Transmission}
\label{subsec:dlpc}
During the OTA-FL broadcasting step, the PS shares the global model parameter with users through downlink channels.
At global iteration $t$, the PS intends to transmit $\x_t^d$ that contains information about global model vector.
Specifically, $\x_t^d$ is:
\begin{equation}
    \x_t^d = \beta_t^d \w_{t}, \label{equ:dltransig}
\end{equation}
where $\beta_t^d$ is the power control (PC) parameter to satisfy the downlink power constraint.
User $i$ receives signal~\eqref{equ:dlrecvsig} and performs the following descaling:
\begin{equation} \label{equ:wt0i}
    \hat{\w}_t =  \frac{1}{\beta_t^d \hh_t^{d,i}} \y_t^{d,i} = \frac{h_t^{d,i}}{\hh_t^{d,i}} \w_{t} + \tilde{\z}_t^{d,i},
\end{equation}
where $\tilde{\z}_t^{d,i}$ is the equivalent scaled downlink noise at user $i$.
Unlike the assumption in~\cite{mao22roar} where all users start with perfect $\w_t$, in this paper, each client $i$ can only initialize its local training with the estimated global model, i.e., $\w^i_{t, 0} = \hat{\w}_t$.
Note that under the imperfect CSI scenario, the estimation not only contains the noise, but also has the misalignment from original signal, which can potentially degrade the learning performance significantly.
We will analyze such impact of noisy downlink with imperfect CSI in Sec.~\ref{sec: conv}.

\vspace{-0.05in}
\subsection{Uplink Transmission}
\label{subsec:uppc}
We consider a dynamic PC scheme following previous works \cite{yang22,mao22,mao22roar}.
Specifically, transmit signal $\x_t^i$ of user $i$ is given by:
\begin{equation}
    \x_t^i = \beta_t^i (\w^i_{t, \tau_t^i} - \w^i_{t, 0}), \label{equ:transig}
\end{equation}
where $\beta_t^i$ is the adaptive PC scaling factor.
After receiving all the local updates, the PS scales the received signal~\eqref{equ:upreceivesig} with $\beta_t^u$.
Hence the PS derives the global model update as: 
\begin{align}
    \w_{t+1} &= \w_{t} + \frac{1}{\beta_t^u}\sum_{i=1}^{m} h_t^{u,i} \x_t^i  + \tilde{\z}_t^u, \label{equ:globalup}
\end{align}
where $\tilde{\z}_t^u$ is the equivalent scaled Gaussian noise with zero mean and variance $\frac{\sigma_{c,u}^2}{(\beta_t^u)^2} \mf{I}_d$.

We discuss a channel inversion policy coupled with dynamic local steps to mitigate the channel fading effect.
In particular, we design the PC factor for user $i$ as:
\begin{equation} \label{equ:betaiimp}
    \beta_t^i = \frac{\beta_t^u \alpha_i}{\tau^i_t \hh_t^{u,i}}.
\end{equation}
where $\tau^i_t$ is the number of local steps at user $i$ in global iteration $t$, and $\hh_t^{u,i}$ is the estimated uplink CSI.
As discussed in Sec.~\ref{subsec:dlpc}, imperfect CSI can cause signal misalignment in aggregation, leading to degradation of the learning performance over iterations.
However, by leveraging dynamic local steps, we can mitigate this impact while also optimizing the local computation resources.
In addition, we set:
\begin{equation} \label{inequ:betai}
    3\eta_t^2 \beta_t^i \tau_t^i G^2 \leq P_t^i,
\end{equation}
where $G$ is defined in Assumption~\ref{a_bounded} in Sec.~\ref{sec: conv}. 
This criterion is used to update the RIS phase matrix in the current round.
Once the phase update finishes, each user decides the number of local steps $\tau_t^i$ to train by satisfying transmit power constraint.
On the other hand, it contributes to the convergence of the proposed algorithm, which will be illustrated in Sec.~\ref{sec: conv}.

\subsection{Phase Design}
\label{subsec:phase}
As shown in Algorithm~\ref{alg:apaf}, the PS updates the RIS phase before FL steps are taken in each global iteration.
Then, the RIS phase shifts remain constant for both uplink and downlink transmissions during the current round.
To determine the RIS phase shifts, we choose the user with the maximum number of local steps in the previous global iteration and design the RIS phase according to its~(\ref{inequ:betai}) with estimated uplink CSI as in~\cite{mao22roar}.
By combining~(\ref{equ:betaiimp}) and~(\ref{inequ:betai}), we get the inequality:
\begin{equation} \label{inequ:phase}
    (\g_t^i)^H \theb_t \geq \frac{3 \eta_t^2 \beta_t^u \alpha_i G^2}{P_t^i} - \hh_{UB,t}^{u,i}, 
\end{equation}
From~\eqref{inequ:phase}, we formulate the phase design problem as:
\begin{equation} \label{prob:phase}
    \begin{aligned} 
        & \mathop{min}\limits_{\theb_t} \quad \|(\g_t^i)^H \theb_t - \frac{3 \eta_t^2 \beta_t^u \alpha_i G^2}{P_t^i} + \hh_{UB,t}^{u,i} \|_2^2  \\
        &\begin{array}{ll}
        s.t. & |\theta_{t,n}|=1 , \quad n=1,...,N.
        \end{array}
    \end{aligned}
\end{equation}

To tackle the non-convex problem~(\ref{prob:phase}), we apply successive convex approximation (SCA) technique to find a stationary solution\cite{scutari2013,mao22papa}.
We first define the objective function as:
\begin{equation} \label{equ:phasefunc}
\begin{array}{ll}
     f(\theb_t) & = || s_t^i - (\g_t^i)^H \theb_t||_2^2   \\
     & = (s_t^i)^* s_t^i - 2 Re \{ \theb_t^H \textbf{v}\} + \theb_t^H \textbf{U} \theb_t ,
\end{array}
\end{equation}
where $s_t^i = \frac{3 \eta^2 \beta_t^u \alpha_i G^2}{P_t^i} - \hh_{UB,t}^i$, $\textbf{v} = s_t^i \g_t^i$, $\textbf{U} = \g_t^i (\g_t^i)^H$.
Next, for ease of computation, we use the equivalent representation for phase element, i.e., $\theta_{n,t} = e^{j \phi_{n,t}}, \phi_{n,t} \in \Rb$.
Since $s_t^i$ is a constant, it is equal to minimize
\begin{equation}
    f_1(\phib_t) = (e^{j\phib_t})^H \textbf{U} e^{j\phib_t} - 2 Re\{(e^{j\phib_t})^H \textbf{v}\},
\end{equation}
where $\phib_t = (\phi_{1,t},...,\phi_{N,t})^T$. Now we introduce SCA method to solve above problem.
We apply the second order Taylor expansion as the surrogate function of $f_1(\phib_t)$ at point $\phib_i^j$:
\begin{equation}
\begin{array}{l}
      g(\phib_t,\phib^j_t) = f_1(\phib^j_t) + \nabla f_1(\phib^j_t)^T (\phib_t - \phib^j_t) + \frac{\lambda}{2} ||\phib_t - \phib^j_t||_2^2,
\end{array}
\end{equation}
where $\nabla f_1(\phib^j_t)$ is the gradient and $\lambda$ is the parameter that satisfies the requirement, i.e, $g(\phib_t,\phib^j_t) \geq f_1(\phib_t) $. 
Then, we set the update step of $\phib_t$ for $J$ iterations as:
\begin{equation}
    \phib_t^{j+1} = \phib^j_t - \frac{\nabla f_1(\phib^j_t)}{\lambda}, j \in [J].\label{equ:phase}
\end{equation}
Finally, we get the desired phase update after SCA process finishes, i.e., $\theb_t= e^{j\phib_t}$.
The server then notifies the RIS controller about the phase information, and RIS changes accordingly.

Algorithm~\ref{alg:apaf} summarizes our proposed joint communication and learning approach.
Different than~\cite{mao22roar}, we introduce PC in the downlink for our practical model.
We integrate RIS phase design into dynamic power control scheme for time-varying channels with imperfect CSI. Our algorithm leverages dynamic local steps to meet communication constraints and contribute to the training process simultaneously.
We demonstrate the robustness of our algorithm in Sec.~\ref{sec: exp}.
\vspace{0.1in}

\section{Convergence Analysis}
\label{sec: conv}
\vspace{-0.05in}
We analyze the convergence behavior of Algorithm~\ref{alg:apaf}.
We first present the assumptions:
\vspace{-0.08in}
\begin{assum}($L$-Smooth) \label{a_smooth}
	 The gradient of loss function satisfies $ \| \nabla F_i(\w_1) - \nabla F_i(\w_2) \| \leq L \| \w_1 - \w_2 \|$, $\forall \w_1, \w_2 \in \mathbb{R}^d$, and $i \in [m]$.
\end{assum}
\vspace{-0.15in}

\begin{assum}(Bounded Variance for Unbiased Local Stochastic Gradients) \label{a_unbias}
	The local stochastic gradient is unbiased and has a bounded variance, i.e.,
	$\mathbb{E} [\nabla F_i(\w, \xi_i)] = \nabla F_i(\w)$, and $\mathbb{E} [\| \nabla F_i(\w, \xi_i) -  \nabla F_i(\w) \|^2] \leq \sigma^2$, $\forall i \in [m]$, where $\xi_i$ is a random sample in $D_i$.
\end{assum}
\vspace{-0.15in}

\begin{assum}(Bounded Stochastic Gradient) \label{a_bounded}
	The norm of the local stochastic gradients is bounded, i.e., $\mathbb{E} [\| \nabla F_i(\w, \xi_i) \|^2] \leq G^2$, $\forall i \in [m]$.
\end{assum}

Then we provide our main convergence result in Theorem~\ref{thm:convergence}.
\vspace{-0.25in}
\begin{restatable}[Convergence Rate] {theorem} {convergence} \label{thm:convergence}
    Define a constant learning rate $\eta_t = \eta  \leq \frac{1}{L}$, and $P_t^i=P_i, \forall t \in [T]$, with Assumptions~\ref{a_smooth}-~\ref{a_bounded}, we have:
    \begin{multline}
        \min_{t \in [T]} \mb{E} \| \nabla F(\w_t) \|^2 \leq \underbrace{\frac{2 \left(F(\w_0) - F(\w_{*}) \right)}{T \eta}}_{\mathrm{optimization \, error}} + \underbrace{\frac{L \sigma_{c,u}^2}{ \eta \beta^2}}_{\substack{\mathrm{channel \,
        noise} \\ \mathrm{error}}}   \nonumber \\
        + \underbrace{ \frac{2 m L^2}{9 \eta^2 G^2}  \sum_{i=1}^m  \frac{(\alpha_i)^2 P_i^2}{(\beta_i^2)} }_{\mathrm{local \, update \, error}} + \underbrace{  L \eta \sigma^2 \frac{1}{T} \sum_{t=0}^{T-1} \sum_{i=1}^{m} \alpha_i^2 \mb{E}_t  \bigg\| \frac{h_t^{u,i}}{\hh_t^{u,i}} \bigg\|^2}_{\mathrm{statistical \, error}} \nonumber \\
        + \underbrace{2 m G^2  \frac{1}{T} \sum_{t=0}^{T-1} \sum_{i=1}^m (\alpha_i)^2 \mb{E}_t  \bigg\| 1 - \frac{h_t^{u,i}}{\hh_t^{u,i}} \bigg\|^2}_{\mathrm{uplink \, channel \, estimation \, error}} \nonumber , \\
        + \underbrace{2 m L^2 \frac{1}{T} \sum_{t=0}^{T-1} \sum_{i=1}^m (\alpha_i)^2 \left( \mb{E}_t \bigg\|1 - \frac{h_{t}^{d,i}}{\hh_{t}^{d,i}} \bigg\|^2 V(t) \right) }_{\mathrm{downlink \, channel \, estimation \, error}} \nonumber ,\\
        + \underbrace{2 m L^2 \frac{1}{T} \sum_{t=0}^{T-1} \sum_{i=1}^m(\alpha_i)^2\frac{\sigma_{c,d}^2}{\| \hh_{t}^{d,i}\|^2 (\beta_t^d)^2}}_{\mathrm{downlink \, channel \, noise \, error}} \nonumber ,
    \end{multline}
    where $\frac{1}{\beta_i^2} = \frac{1}{T} \sum_{t=0}^{T-1} \frac{1}{(\beta_t^i)^2}$, $\frac{1}{\bar{\beta}^2} = \frac{1}{T} \sum_{t=0}^{T-1} \frac{1}{(\beta_t^u)^2}$, and
    \begin{equation*}
        V(t) \triangleq 2 \|\w_0\|^2 + \sum_{l=0}^{t-1} \frac{\sigma_{c,u}^2}{(\beta_l^u)^2} + 2 t m G^2 \eta^2\sum_{l=0}^{t-1}  \sum_{i=1}^m \alpha_i^2 \left( \frac{h_l^{u,i}}{\hh_l^{u,i}}\right)^2.
    \end{equation*}
\end{restatable}
\vspace{-0.01in}
\begin{proof}[Proof Highlights] 
Early steps of the proof are similar to~\cite{mao22, mao22roar}.
We start with one-step loss function descent from $L$-smooth in Assumption~\ref{a_smooth},
then insert the global model update~\eqref{equ:globalup} and expand every term to find the corresponding upper bound employing Cauchy-Schwartz inequality with Assumptions~\ref{a_smooth}-\ref{a_bounded}.
In this procedure, we decouple the channel estimation, channel noise, and learning parameters because they are independent.
Note that the presence of noisy downlink results in a noisy starting point for each user in every training round.
As such, the key internal step is
\vspace{-0.01in}
\begin{align}
    &\mb{E}_t \|\w_t - \w^i_{t, k} \|^2  = \mb{E}_t \|\w_t - \w_{t,0}^i + \w_{t,0}^i - \w^i_{t, k} \|^2  \\
    & \leq \mb{E}_t \bigg\|1 - \frac{h_{t}^{d,i}}{\hh_{t}^{d,i}} \bigg\|^2 \mb{E}_t \|\w_t\|^2 + \frac{\sigma_{c,d}^2}{(\beta_t^d)^2\| \hh_{t}^{d,i}\|^2} + \eta_t^2 k^2 G^2
    \label{equ:noisydl}
\end{align}
We continue to find the bound for $\mb{E}_t \|\w_t\|^2$ as $V(t)$. 
Finally, by applying constraint~\eqref{inequ:betai} and Jensen's inequality, we obtain the convergence rate.
\end{proof}

Theorem~\ref{thm:convergence} shows that there are seven resources of error on the convergence upper bound: 1) the FL optimization error; 2) uplink channel noise error; 3) the local update error; 4) the learning statistical error; 5) uplink channel estimation error; 6) downlink channel estimation error; 7) downlink channel noise error.
The learning error types are coupled with the communication error types due to the joint communication and learning design.
In this paper, we consider a realistic RIS-assisted communication model with noisy downlink and uplink.
As a result, we get two additional errors related to noisy downlink compared to~\cite{mao22roar} that assumes error-free downlink.
Moreover, we assume that users only have imperfect CSI, which inevitably leads to two channel estimation errors.
Note that in the perfect CSI case, i.e., $h_t^i = \hh_t^i$, these two estimation errors will diminish, which matches the literature that fading channels can be fully alleviated by power control.
In particular, because of the RIS assistance, our design can still achieve excellent learning performance when the direct links are weak.

To bound the statistical error and channel estimation error types, we use the Taylor expansion to analyze the channel estimation error similar to \cite{zhu2020one}: $\frac{h_t^i}{\hh_t^i} = \frac{1}{1 + \frac{\Delta_t^i}{h_t^i}}  = 1 - \frac{\Delta_t^i}{h_t^i} + \mc{O}( (\frac{\Delta_t^i}{h_t^i})^2)$.
This applies to both uplink and downlink.
By ignoring the higher order terms, we get the Corollary 1:
\begin{restatable}{corollary} {convergence_rate} \label{cor:convergence}
Let $|\Delta_t| \ll |h_t|, \forall t\in[T]$, $h_{UB,m} = \mathop{min}\limits_{t \in [T], i \in[m]}\{|h_{UB,t}^{i}|\}$, $h_{UR,a} = \mathop{max}\limits_{t \in [T], i \in[m], j \in [N]}\{|h_{UB,t,j}^{i}|\}$, $h_{RB,a} = \mathop{max}\limits_{t \in [T], j \in [N]}\{|h_{RB,t,j}|\}$, $g_m=\mathop{min}\limits_{t \in [T], i \in[m], j \in [N]}\{|g_{t,j}^i|\}$ for both downlink and uplink, the convergence rate of Algorithm~\ref{alg:apaf} is bounded. The statistical error and channel estimation errors are bounded by:
\begin{multline}
        L \eta \sigma^2 \frac{1}{T} \sum_{t=0}^{T-1} \sum_{i=1}^{m} \alpha_i^2 \mb{E}_t  \bigg\| \frac{h_t^{u,i}}{\hh_t^{u,i}} \bigg\|^2  \leq 
        L \eta \sigma^2 \sum_{i=1}^{m} \alpha_i^2 \left( 1 + C\right), \nonumber \\
        2 m G^2  \frac{1}{T} \sum_{t=0}^{T-1} \sum_{i=1}^m (\alpha_i)^2 \mb{E}_t  \bigg\| 1 - \frac{h_t^{u,i}}{\hh_t^{u,i}} \bigg\|^2 \nonumber \leq
        2 m G^2 \sum_{i=1}^m (\alpha_i)^2 C, \nonumber \\
        2 m L^2 \frac{1}{T} \sum_{t=0}^{T-1} \sum_{i=1}^m (\alpha_i)^2 \left( \mb{E}_t \bigg\|1 - \frac{h_{t}^{d,i}}{\hh_{t}^{d,i}} \bigg\|^2 V(t) \right)  \quad \quad \quad \quad \\ \leq 2 m L^2 \sum_{i=1}^m(\alpha_i)^2CV(T),  
\end{multline}
where $C=\frac{\sigt_h^2 (1+N^2(h_{UR,a}^2+h_{RB,a}^2+\sigt_h^2))}{(h_{UB,m})^2}$. 
\end{restatable}

The number of RIS elements $N$ influences the convergence bound as illustrated in Corollary~\ref{cor:convergence}.
With the increasing $N$, the channel estimation error increases because of more links in the system, thus the corresponding convergence bounds also increase.
However, this error type is not dominant, $N$ is coupled with a small estimation variance.
Having more RIS elements results in better channel conditions, which in turn leads to smaller local update and downlink channel noise errors. This ultimately enhances the learning performance.
We also demonstrate such impact in Sec.~\ref{sec: exp}.

\section{Numerical Results} \label{sec: exp}
We consider an image classification task on the MNIST dataset~\cite{lecun1998gradient} by training a logistic regression model.
We assume an extreme heterogeneous non-i.i.d. case, i.e., each local dataset only contains {\it one} class data points.

We set up an RIS-assisted multiuser communication system.
Specifically, the system consists of $m=10$ users and a single RIS with $N=45$ elements.
Following~\cite{liu2021risfl}, we consider a 3D coordinate system, where the location of BS is (-50,0,10) meters and RIS is deployed at (0,0,10) meters.
We distribute the users uniformly in the x-y plane within the range of [-20,0] meters in the x dimension and [-30,30] meters in the y dimension.
The channel model follows i.i.d. Gaussian distribution with scaling of the square root of path loss.
We consider the path loss model in~\cite{tang2020ris}.
The user-PS direct link model is  $G_{PS}G_{U}\left(\frac{3*10^8 m/s}{4 \pi f_c d_{UP}} \right)^{PL}$, where we set path loss exponent $PL=4$ to simulate weak direct link, $G_{PS}=5$dBi, $G_U=0$dBi are antenna gains, $d_{UP}$ is the user-PS distance, and $f_c=915$MHz is the carrier frequency.
The RIS assisted link model is $G_{PS}G_{U} G_{RIS} \frac{N^2 d_x d_y ((3*10^8 m/s)/f_c)^2 }{64 \pi^3 d_{RP}^2 d_{UR}^2}$, where $d_x=d_y=(3*10^7 m/s)/f_c$ are the dimensions of an RIS element, $G_{RIS}=5$dBi is the RIS antenna gain, and $d_{RP}, d_{UR}$ are RIS-PS distance, user-RIS distance, respectively.
To simulate an imperfect CSI scenario, we define the channel estimation error of each path as i.i.d. Gaussian random variable with variance $\sigt_h^2= 0.1 \sigma_{c,u}^2$.
The transmit uplink SNR is set to \unit[$20$]{dB}, while the transmit downlink SNR is set to \unit[$30$]{dB} since the server is usually power-sufficient.

We compare our proposed algorithm with two baselines:
\begin{list}{\labelitemi}{\leftmargin=1em \itemindent=-0.5em \itemsep=.2em}
\item Baseline 1: Algorithm~\ref{alg:apaf} with noiseless downlink (\alg~\cite{mao22roar}).
\item Baseline 2: Algorithm from~\cite{liu2021risfl}.
\end{list}

\begin{figure}[t] 
    \centering
    \includegraphics[scale=0.5]{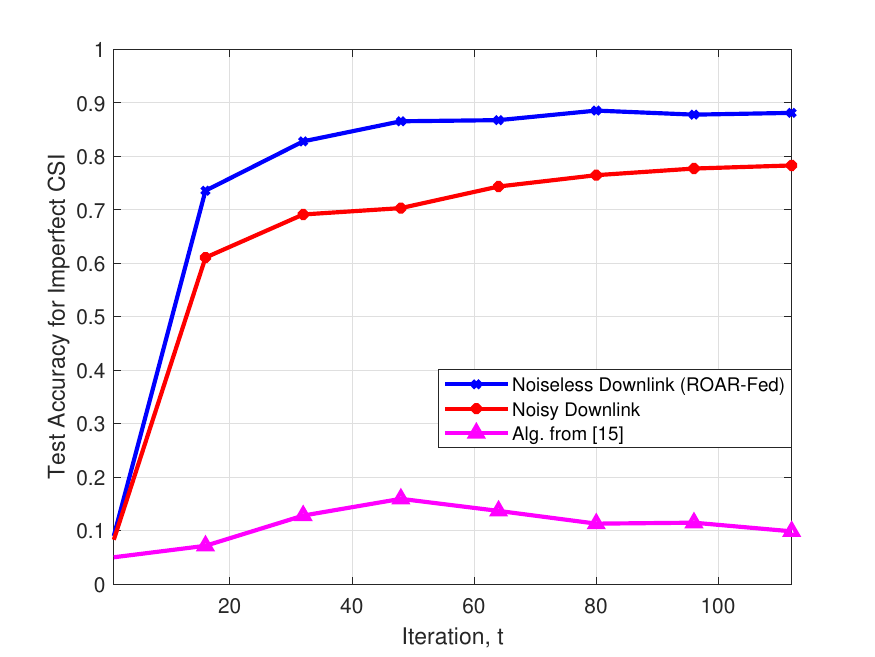}
    \caption{Test accuracy.}
    \label{fig:result1}
\end{figure}
\vspace{-0.04in}
Fig.~\ref{fig:result1} illustrates the results on the test accuracy of the proposed algorithm and above benchmarks versus the number of global iterations.
We can see that, even with the noisy downlink and imperfect CSI, the proposed algorithm achieves excellent learning performance.
Compared with the noiseless downlink case, the performance of the proposed algorithm degrades as expected, which clearly demonstrates the nontrivial impact of the noisy downlink. That is, noisy downlink degrades learning performance, which matches our theoretical analysis.
Note that the algorithm from~\cite{liu2021risfl} fails to converge in this realistic time-varying noisy channels with imperfect CSI and heterogeneous non-i.i.d. data setting.
Our algorithm outperforms it and is convergent, which further verifies the effectiveness and robustness of our joint adaptive learning and communication design.

\begin{figure}[t] 
    \centering
    \includegraphics[scale=0.5]{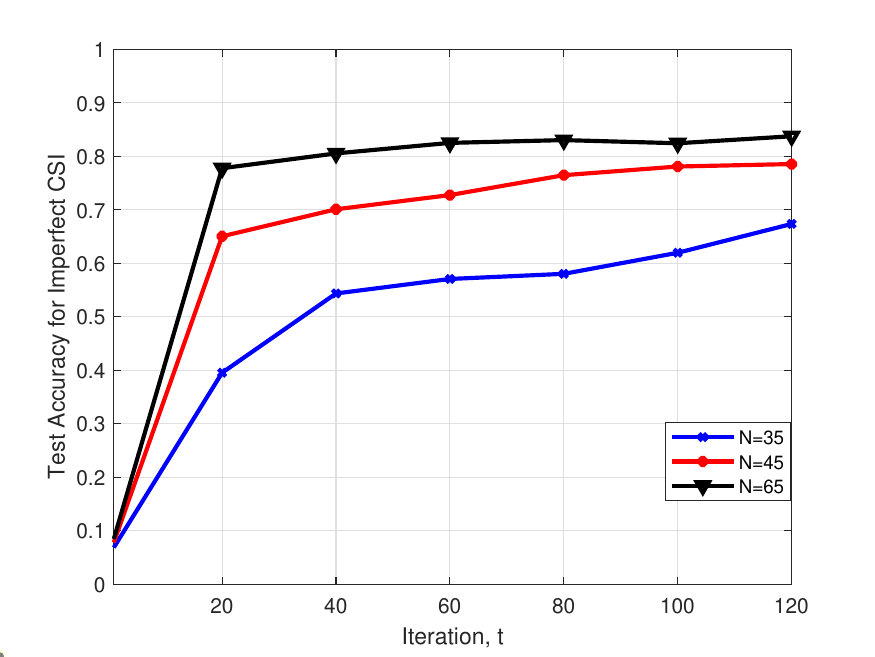}
    \caption{Test accuracy versus the number of RIS elements.}
    \label{fig:result2}
\end{figure}
\vspace{-0.04in}
Fig.~\ref{fig:result2} shows the impact of the number of reflecting elements at the RIS on the test accuracy.
When $N$ increases, the convergence speed and test accuracy increase together, which matches our theoretical analysis in Corollary 1.
It is worth mentioning that the larger $N$ means the better propagation environment that RIS facilitates for both uplink and downlink, resulting in less local update error which is more dominant than the growing channel estimation errors. 

\vspace{-0.05in}
\section{Conclusion} 
\label{sec: conclusion}
In this paper, we have investigated an RIS-assisted heterogeneous network for federated edge learning.
We have proposed a new adaptive cross-layer approach that jointly optimizes communication and computation resources over uplink and downlink noisy communications.
In particular, the number of local steps in local training is coupled with transmit power and RIS configuration under imperfect CSI case for a general non-convex learning objective.
We have analyzed the convergence of the proposed algorithm and revealed the impact of RIS. 
We have shown the effectiveness and robustness of our approach, which outperforms the existing RIS-assisted OTA-FL design under heterogeneous data and imperfect CSI.

\bibliographystyle{IEEEtran}{}
\bibliography{BIB/Optimization,BIB/Relay, BIB/RIS, BIB/ImperfectCSI, BIB/Yener, BIB/RISFL, BIB/Experiments, BIB/Introduction}


\end{document}

%% file: symbols_commands.tex
\newcommand{\Cb}{\mathbb{C}}

\newcommand{\g}{\mathbf{g}}
\newcommand{\Hb}{\mathbf{H}}
\newcommand{\h}{\mathbf{h}}
\newcommand{\hh}{\widehat{h}}

\newcommand{\Rb}{\mathbb{R}}

\newcommand{\sigt}{\widetilde{\sigma}}

\newcommand{\w}{\mathbf{w}}

\newcommand{\x}{\mathbf{x}}

\newcommand{\y}{\mathbf{y}}
\newcommand{\z}{\mathbf{z}}

\newcommand{\mc}[1]{\mathcal{#1}}
\newcommand{\mb}[1]{\mathbb{#1}}
\newcommand{\mf}[1]{\mathbf{#1}}

\newtheorem{assum}{Assumption}

\newcommand{\The}{\mathbf{\Theta}}
\newcommand{\theb}{\boldsymbol{\theta}}
\newcommand{\phib}{\boldsymbol{\phi}}



%% file: Abstract.tex

\begin{abstract}
Over-the-air federated learning (OTA-FL) exploits the inherent superposition property of wireless channels to integrate the communication and model aggregation.
Though a naturally promising framework for wireless federated learning, it requires care to mitigate physical layer impairments.
In this work, we consider a heterogeneous edge-intelligent network with different edge device resources and non-i.i.d. user dataset distributions, under a general non-convex learning objective.
We leverage the Reconfigurable Intelligent Surface (RIS) technology to augment OTA-FL system over simultaneous time varying uplink and downlink noisy communication channels under imperfect CSI scenario.
We propose a cross-layer algorithm that jointly optimizes RIS configuration, communication and computation resources in this general realistic setting.
Specifically, we design dynamic local update steps in conjunction with RIS phase shifts and transmission power to boost learning performance.
We present a convergence analysis of the proposed algorithm, and show that it outperforms the existing unified approach under heterogeneous system and imperfect CSI in numerical results.

\end{abstract}